\documentclass{article}
\usepackage{spconf,amsmath,graphicx, mathtools}
\usepackage{amssymb}
\usepackage{booktabs}
\usepackage[english]{babel}

\newcommand{\ours}{COLA }
\newcommand{\norm}[1]{\left\lVert#1\right\rVert}
\newcommand{\efficientnet}{EfficientNet-B0~}
\usepackage{xcolor}
\usepackage{makecell}

\def\x{{\mathbf x}}

\def\encoder{{f}}
\def\projection{{g}}
\def\embedding{{h}}

\title{Contrastive learning of general-purpose audio representations}
\name{Aaqib Saeed$^1$\sthanks{This work was conducted while interning at Google.}, David Grangier$^2$, Neil Zeghidour$^2$}
\address{$^1$Eindhoven University of Technology, Eindhoven, The Netherlands \\ 
$^2$Google Research, Paris, France
}

\begin{document}
\maketitle
\begin{abstract}
We introduce COLA, a self-supervised pre-training approach for learning a general-purpose representation of audio. Our approach is based on contrastive learning: it learns a representation which assigns high similarity to audio segments extracted from the same recording while assigning lower similarity to segments from different recordings. We build on top of recent advances in contrastive learning for computer vision and reinforcement learning to design a lightweight, easy-to-implement self-supervised model of audio. We pre-train embeddings on the large-scale Audioset database and transfer these representations to 9 diverse classification tasks, including speech, music, animal sounds, and acoustic scenes. We show that despite its simplicity, our method significantly outperforms previous self-supervised systems. We furthermore conduct ablation studies to identify key design choices and release a library\footnote{\scriptsize{\url{https://github.com/google-research/google-research/tree/master/cola}}} to pre-train and fine-tune COLA models.
\end{abstract}
\section{Introduction}
\label{sec:intro}
\begin{figure*}[t]
\includegraphics[width=\textwidth]{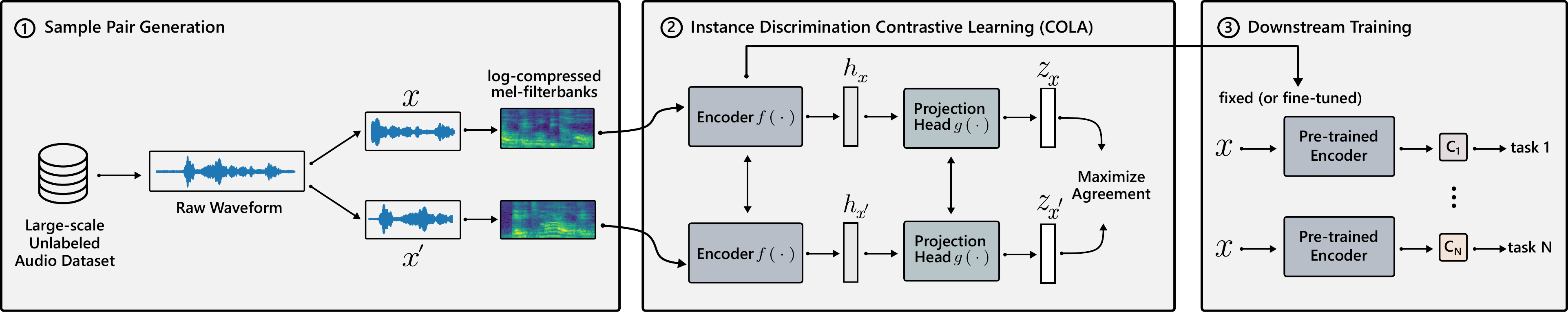}
\caption{Overview of the contrastive self-supervised learning for audio.}
\label{fig:overview}
\end{figure*}

Self-supervised pre-training has recently emerged as a successful technique to leverage unlabeled data to learn representations beneficial to supervised problems. This success spans a wide range of tasks and modalities~\cite{grill2020bootstrap, oord2018representation, devlin2018bert, owens2018audio}.
Among these methods, Discriminative Pre-Training (DPT) is particularly effective. This approach learns a representation from pairs of \textit{similar} inputs from unlabeled data, exploiting e.g. temporal consistency~\cite{kawakami2020learning, owens2018audio, jansen2018unsupervised} or data augmentation~\cite{chen2020simple} and trains a model to recognize \textit{similar} elements among negative distractors. In contrast with generative encoder-decoder approaches \cite{chung2016audio, plchot2016audio, meyer2017unsupervised, wan2017google, pascual2019learning}, DPT is computationally efficient as it avoids input reconstruction entirely. 

Amidst DPT models for audio, ~\cite{jansen2018unsupervised} used a metric learning approach with a triplet loss to minimize the distance between embeddings of anchor and positive pairs and maximize it among the negatives. The instance generation is achieved through noise injection, shifting along time-frequency dimensions, and extracting samples in temporally close neighborhoods. Along similar lines,~\cite{shor2020towards} proposed a benchmark for comparing speech representations on non-semantic tasks. Through utilizing a triplet loss as an unsupervised objective with a subset of AudioSet~\cite{gemmeke2017audio} for model training, they showed improved performance on several downstream speech classification tasks. Inspired from seminal work in NLP~\cite{mikolov2013efficient}, the work in ~\cite{tagliasacchi2019self} adopted a similar approach to learn audio representations (i.e. \textsc{Audio2Vec}) along with another ``pretext'' task of estimating temporal distance between audio segments. The pre-trained models are tested on several downstream tasks, from speaker identification to music recognition.
Despite recent progress, most work on learning representations of audio focuses on speech tasks~\cite{baevski2019vq,riviere2020unsupervised,baevski2020wav2vec} (with the exception of~\cite{jansen2018unsupervised,tagliasacchi2019self}) and ignores other audio tasks such as acoustic scene detection or animal vocalizations. Moreover, triplet-based objectives heavily rely on the mining of negative samples, and the quality of learned features can vary significantly with the sample generation scheme.

In this work, we propose \ours(COntrastive Learning for Audio), a simple contrastive learning framework to learn general-purpose representations of sounds beyond speech. We build upon recent advances in contrastive learning \cite{oord2018representation} for computer vision ($\textsc{SimCLR}$~\cite{chen2020simple}, $\textsc{MoCo}$~\cite{he2020momentum}) and reinforcement learning ($\textsc{CURL}$~\cite{srinivas2020curl}). We generate similar pairs by simply sampling segments from the same audio clip, which avoids exploring augmentation strategies entirely unlike $\textsc{SimCLR}$, $\textsc{MoCo}$, $\textsc{CURL}$ and others \cite{ kharitonov2020data}. Our dissimilar pairs simply associate segments from different clips in the same batch, which does not require maintaining a memory bank of distractors as in $\textsc{MoCo}$. Our approach allows us to consider a large number of negatives for each positive pair in the loss function and bypass the need for a careful choice of negative examples, unlike triplet-based approaches~\cite{jansen2018unsupervised, shor2020towards}. \ours is also different from CPC~\cite{oord2018representation} as it does not predict future latent representations from past ones.

We demonstrate the effectiveness of \ours over challenging and diverse downstream tasks, including speech, music, acoustic scenes, and animal sounds. After pre-training on the large-scale AudioSet database~\cite{gemmeke2017audio}, we show that a linear classifier trained over a \ours embedding gets close to the performance of a fully-supervised in-domain convolutional network and exceeds it when using fine-tuning. Moreover, our system outperforms previous unsupervised approaches on most downstream tasks. These experiments demonstrate that \ours offers a simple, easy-to-implement method to learn general-purpose audio representations without supervision.

\section{Method}
\label{sec:method}
We learn general-purpose audio representations from unlabeled data by pre-training a neural network with a contrastive loss function. 
Our objective function maximizes an agreement between the latent embedding of segments extracted from the same audio clip while using different audio clips as negative classes, as shown in Figure~\ref{fig:overview}. This objective pre-trains a convolutional feature extractor on unlabeled audio data. After pre-training, we combine our feature extractor with an additional classification layer for solving various audio understanding tasks across several datasets.

Contrastive learning extracts a latent space in which the similarity between an anchor example and a related example should be greater than the similarity between the same anchor and unrelated examples. In our case, an anchor and its corresponding positive are audio segments from the same clip. This contrasts with approaches that generate positives as perturbations of the anchor~\cite{wu2018unsupervised,kharitonov2020data}. For negative examples, we take segments from different audio clips in the current training batch. This strategy allows to consider a large number of negatives and is efficient since batch examples are used both as positives and negatives without additional computation.

\ours computes the similarity between audio segments in two steps. First, an encoder $\encoder$ maps a log-compressed mel-filterbanks $\x \in \mathbb{R}^{N \times T}$, with $N$ and $T$ the number of frequency bins and time frames respectively, into a latent representation $h = \encoder(x) \in \mathbb{R}^{d}$. This is the representation that we will transfer to downstream classification, after pre-training. Then, a shallow neural network $\projection$ maps $h$ onto a space $z = \projection(h)$, where bilinear comparisons are performed. If we denote with $W$ the bilinear parameters, the similarity between two segments ($x$, $x'$) is, therefore:
\begin{equation}
{\rm s}(x, x') = {\projection}({\encoder}(x))^\top ~W~ {\projection}({\encoder}(x')).
\end{equation}
Bilinear similarity has been used in the past~\cite{oord2018representation} but is less common than cosine similarity, e.g. \textsc{SimCLR} and \textsc{MoCo}. In Section \ref{sec:experiments}, we perform an ablation study on the choice of similarity measure. Table \ref{tab:ablation_similarity} shows that a bilinear similarity outperforms a simple cosine similarity ($\frac{{\projection}({\encoder}(x))^\top \cdot {\projection}({\encoder}(x'))}{\norm{{\projection}({\encoder}(x))}\norm{{\projection}({\encoder}(x'))}}$) on all downstream tasks. In the rest of this paper, we use this method when not stated otherwise.

As an objective function, we rely on multi-class cross entropy applied to similarities, i.e.
\begin{equation}
    \mathcal{L} = -\log \frac{\exp\left( {\rm s}(x, x^+)\right)}
   {\sum\limits_{\mathclap{x^- \in {\cal X^-}(x)\cup\{x^+\}}} \exp\left( {\rm s}(x, x^-)\right) } 
\end{equation}
where $x^+$ is the positive associated to anchor $x$, while ${\cal X^-}(x)$ refers to the set of negative distractors.
This loss, unlike the triplet loss~\cite{wang2015unsupervised}, leverages multiple distractors at a time. %

As mentioned earlier, we train our model with positive segment pairs sampled from the same audio clip. For each pair, we use one segment as the anchor and the other element as the positive. Positive segments are used as negatives for all other anchors in the batch. This strategy is more efficient than keeping a memory bank of negatives~\cite{wu2018unsupervised, he2020momentum} since the representation of an example is paired with every anchor in the batch either as a positive or as a negative segment. In particular, we experiment with batch sizes varying from $256$ to $2048$, as shown in Table \ref{tab:ablation_batch_size}. A large batch size allows the model to see many negative samples per anchor and helps accuracy on end tasks. It is important to note that we sample segment pairs on-the-fly and reshuffle the data at each training epoch to maximize the diversity of positive and negative pairs seen during training. The sample generation procedure is illustrated in Figure~\ref{fig:overview}. 

\section{Experiments}
\label{sec:experiments}
We evaluate our method by pre-training \ours embeddings on a large-scale audio dataset and then transferring it to downstream tasks in the following ways: 1) training a linear classifier on top of a frozen embedding, used as a feature extractor and 2) fine-tuning the entire network on the end-task. Importantly, we assess the performance on several diverse datasets to determine the transferability of learned representations across audio domains and recording conditions. %

\subsection{Datasets and Tasks}
We pre-train \ours embeddings on the diverse, large-scale Audioset~\cite{gemmeke2017audio}. It contains $2$ millions excerpts of $10$ seconds audio from YouTube videos that are annotated in a multi-label fashion with over $500$ classes. This dataset has been used by \cite{tagliasacchi2019self, tagliasacchi2020pre,shor2020towards} for self-supervised pre-training. Since our method is self-supervised, we never use Audioset labels. As described earlier, we randomly sample audio clips to generate examples. Likewise, for the extraction of anchors and positives, segments of audio are selected uniformly at random inside a sequence.

We perform downstream evaluation on a variety of tasks, including both speech and non-speech. To allow for comparison with previous methods, we rely on datasets that have been previously used by \cite{tagliasacchi2019self, tagliasacchi2020pre, shor2020towards}. For speaker identification, we use a $100$-hours subset of LibriSpeech (LBS)~\cite{panayotov2015librispeech} that contains audio of books read by $251$ speakers, as well as the Voxceleb ~\cite{nagrani2017voxceleb} subset used in \cite{shor2020towards}, with $1,251$ speakers. For keyword spotting, we use Speech Commands (SPC)~\cite{warden2018speech} V$1$ and V$2$ to recognize $11$ and $35$ spoken commands (classes) from one second of audio, respectively. For acoustic scene classification, we use TUT Urban Acoustic Scenes $2018$ (TUT)~\cite{heittola_2018}, consisting of labeled audio segments from $10$ different acoustic scenes. For animal vocalizations, we use the Bird Song Detection (BSD) dataset~\cite{stowell2019automatic} from DCASE $2018$ Challenge to solve a binary classification problem. For music recognition, we use MUSAN~\cite{snyder2015musan} that differentiates audio samples across $3$ classes (speech, music and noise), as well as the NSynth dataset~\cite{engel2017neural} of musical notes, labeled with the family of the instrument (11 classes). For language identification, we use the Voxforge dataset~\cite{maclean2018voxforge} to categorize audio clips into six classes based on the spoken language.

\begin{table}[!t]
\centering
\caption{Test accuracy (\%) on downstream tasks.}
\vspace{0.1cm}
\footnotesize
\setlength\tabcolsep{3pt}
\label{tab:results_baseline}
\begin{tabular}{@{}lcccc@{}}
\toprule
\textbf{}                        & \textbf{Random} & \textbf{Supervised} & \multicolumn{2}{c}{\textbf{\ours}} \\
\textbf{Task}                    & \textbf{Init.}  & \textbf{}           & \textbf{Frozen}  & \textbf{Fine-tuned}  \\ \midrule
\textbf{Speaker Id. (LBS)}       & 0.4             & \textbf{100.0}      & \textbf{100.0}   & \textbf{100.0}       \\
\textbf{Speech commands (V1)}    & 62.9            & 97.2                & 71.7             & \textbf{98.1}        \\
\textbf{Speech commands (V2)}    & 4.0             & 94.3                & 62.4             & \textbf{95.5}        \\
\textbf{Acoustic scenes}         & 8.6             & 98.2                & 94.1             & \textbf{99.2}        \\
\textbf{Speaker Id. (Voxceleb)}  & 0.0             & 31.7                & 29.9             & \textbf{37.7}        \\
\textbf{Birdsong detection}      & 49.6            & 79.4                & 77.0             & \textbf{80.2}        \\
\textbf{Music, Speech and Noise} & 56.8            & 99.3                & 99.1             & \textbf{99.4}        \\
\textbf{Language Id.}            & 59.1            & \textbf{85.0}       & 71.3             & 82.9                 \\
\textbf{Music instrument}        & 20.8            & 70.7                & 63.4             & \textbf{73.0}        \\ \midrule
\textbf{Average}                 & 29.1            & 83.9                & 74.3             & \textbf{85.1}                 \\ \bottomrule
\end{tabular}
\end{table}

\begin{table*}[!htbp]
\centering
\caption{Test accuracy (\%) of a linear classifier trained on top of \ours embeddings or baseline pre-trained representations.}
\vspace{0.1cm}
\footnotesize
\label{tab:results_sota}
\begin{tabular}{@{}lcccccc@{}}
\toprule
 &
\textbf{CBoW~\cite{tagliasacchi2019self, tagliasacchi2020pre}} &
\textbf{SG~\cite{tagliasacchi2019self, tagliasacchi2020pre}} &
\textbf{TemporalGap~\cite{tagliasacchi2019self, tagliasacchi2020pre}} &
\textbf{Triplet Loss~\cite{tagliasacchi2019self, tagliasacchi2020pre}} &
\textbf{TRILL~\cite{shor2020towards}} &
\textbf{\ours} \\ \midrule
\textbf{Speaker Id. (LBS)}       & 99.0  & \textbf{100.0} & 97.0 & \textbf{100.0} & -             & \textbf{100.0} \\
\textbf{Speech commands (V2)}    & 30.0  & 28.0           & 23.0 & 18.0           & -             & \textbf{62.4}  \\
\textbf{Acoustic scenes}         & 66.0  & 67.0           & 63.0 & 73.0           & -             & \textbf{94.0}  \\
\textbf{Birdsong detection}      & 71.0  & 69.0           & 71.0 & 73.0           & -             & \textbf{77.0}  \\
\textbf{Music, Speech and Noise} & 98.0  & 98.0           & 97.0 & 97.0           & -             & \textbf{99.1}  \\
\textbf{Music instrument}        & 33.5  & 34.4           & 35.1 & 25.7           & -             & \textbf{63.4}  \\
\textbf{Speech commands (V1)}    & -     & -              & -    & -              & \textbf{74.0} & 71.7           \\
\textbf{Speaker Id. (Voxceleb)}  & -     & -              & -    & -              & 17.7          & \textbf{29.9}  \\
\textbf{Language Id.}            & -     & -              & -    & -              & \textbf{88.1} & 71.3           \\ \midrule
\textbf{Average (TRILL tasks)}   & -     & -              & -    & -              & {\bf 59.9}    & 57.6           \\
\textbf{Average (non-TRILL)}     & 66.25 & 66.0           & 64.3 & 64.4           & -             & \textbf{82.5}  \\ \bottomrule
\end{tabular}

\end{table*}

\begin{table}[!htbp]
\centering
\caption{Test accuracy (\%) with different similarity functions.}
\vspace{0.1cm}

\footnotesize
\label{tab:ablation_similarity}
\begin{tabular}{@{}lcc@{}}
\toprule
& \textbf{Cosine  Similarity} & \textbf{Bilinear Similarity} \\ \midrule
\textbf{Speaker Id. (LBS)}       & 99.9                        & \textbf{100.0}                     \\
\textbf{Speech commands (V1)}    & 64.5                        & \textbf{71.7}                      \\
\textbf{Speech commands (V2)}    & 42.4                        & \textbf{62.4}                     \\
\textbf{Acoustic scenes}         & 87.5                        & \textbf{94.1}                      \\
\textbf{Speaker Id. (Voxceleb)}  & 15.2                        & \textbf{29.9}                      \\
\textbf{Birdsong detection}      & 76.5                        & \textbf{77.0}                      \\
\textbf{Music, Speech and Noise} & 99.0                        & \textbf{99.1}                      \\
\textbf{Language Id.}            & 62.3                        & \textbf{71.3}                      \\
\textbf{Music instrument}        & 58.3                        & \textbf{63.4}                      \\ \midrule
\textbf{Average}                 & 67.2                        & \textbf{74.3}                      \\ \bottomrule
\end{tabular}
\end{table}

\subsection{Model Architecture and Implementation Details}
Given an audio input sequence, we extract log-compressed mel-filterbanks with a window size of $25$ ms, a hop size of $10$ ms, and $N=64$ mel-spaced frequency bins in the range $60$–$7800$ Hz for $T=96$ frames, corresponding to $960$ ms. These features are passed through an encoder $\encoder$ based on \efficientnet \cite{tan2019efficientnet}, a lightweight and highly scalable convolutional neural network. Even though \efficientnet has been originally proposed for computer vision, the 2D structure of mel-filterbanks allows using this architecture without any adjustment. %
We apply a global max-pooling to the last layer of the encoder to get an embedding $\embedding$ of size $1280$. During pre-training, we pass $\embedding$ through the projection head $\projection$, which contains a fully-connected layer with $512$ units followed by a Layer Normalization~\cite{ba2016layer} and a $\tanh$ activation.  We discard the projection head for the downstream tasks and train a linear classifier on top of the encoder directly. We pre-train all our models with ADAM~\cite{kingma2014adam} and a learning rate of $10^{-4}$, for $500$ epochs. We explore the impact of the batch size and report the results in Table \ref{tab:ablation_batch_size}. We train the downstream classifiers with a batch size of $64$ and a learning rate of $10^{-3}$, on randomly selected $960$ms segments, as for pre-training. However, we evaluate downstream classifiers on entire sequences using the following procedure: we split the sequence into non-overlapping $960$ms segments, pass them through the encoder and linear classifier, and average the predictions.

\subsection{Results}
\label{sec:results}

Table \ref{tab:results_baseline} reports the accuracy on the $9$ downstream datasets. We compare our approach against multiple baselines: a linear classifier trained on a randomly initialized fixed encoder and a fully-supervised model trained directly on downstream datasets which indicates the performance achievable with \efficientnet on these datasets. First, we evaluate pre-trained \ours embeddings with a linear classifier on top of frozen representations, following the same procedure as ~\cite{tagliasacchi2019self, oord2018representation, chen2020simple, he2020momentum}. This outperforms drastically the performance of a linear classifier trained on a random embedding ($74.3\%$ against $29.1\%$ on average), showing that the encoder has learned useful representations. This is remarkable as we pre-train a single \ours embedding, which performs well across many tasks. Next, we also use a pre-trained \ours as initialization and fine-tune one model per downstream task. Table~\ref{tab:results_baseline} shows that on all tasks but language identification, initializing a supervised model with \ours improves the performance over training from scratch ($85.1\%$ against $83.9\%$ on average), which demonstrates the benefits of transferring \ours representations even in a fully supervised setting. 

We then compare \ours to prior self-supervised methods proposed in~\cite{tagliasacchi2019self, tagliasacchi2020pre}, including a standard triplet loss, \textsc{Audio2Vec} (CBoW and SG) and temporal gap prediction models. Here, the CBoW and SG are generative models inspired from \textsc{Word2Vec}, trained to reconstruct a randomly selected temporal slice of log-mel spectrograms given the rest or vice versa. Likewise, TemporalGap trains a model to predict the temporal distance between two pairs of audio segments. Table~\ref{tab:results_sota} shows that \ours embeddings consistently outperform all these methods. In particular, on acoustic scene classification, we obtain a competitive accuracy of $94$\% compared to $73$\% achieved with a triplet loss in~\cite{tagliasacchi2019self}. We also considerably improve the performance on speech commands and musical instrument classification by an absolute $~30\%$ margin on both tasks. We also compare with the recent self-supervised learning framework TRILL~\cite{shor2020towards} on three speech-related tasks, benchmarking against TRILL-19 (the best self-supervised system of \cite{shor2020towards}). Our general-purpose \ours embeddings are competitive with TRILL, despite the fact that TRILL is pre-trained specifically on the part of Audioset that contains speech, and is evaluated only across speech tasks, while we train and evaluate \ours across speech, music, acoustic scenes, and animal sounds.

To investigate the role of the similarity measure in the quality of learned representations, we perform an ablation study to compare model pre-training with cosine and bilinear similarity. With the cosine similarity, we use a temperature $\tau=0.2$ to normalize the scores before computing the loss. Table~\ref{tab:ablation_similarity} reports the results obtained on downstream classifiers using encoders pre-trained with each of the similarity estimation techniques. We observe that the best results are obtained using bilinear similarity in all cases. We also conduct an experiment to measure the impact of pre-training batch size, as larger batch sizes result in more negative samples and facilitate convergence~\cite{chen2020simple}. Table~\ref{tab:ablation_batch_size} shows that, on average, a batch size as large as $1024$ provides better representations compared to smaller ones. However, increasing the batch size up to $2048$ worsens the performance in most cases. 

\begin{table}[t]
\centering
\caption{Impact of pre-training batch size on downstream test accuracy (\%), using a bilinear similarity.}
\vspace{0.1cm}

\footnotesize
\label{tab:ablation_batch_size}
\begin{tabular}{@{}lcccc@{}}
\toprule
& \textbf{256} & \textbf{512}  & \textbf{1024}  & \textbf{2048} \\ \midrule
\textbf{Speaker Id. (LBS)}       & 99.9         & 99.9          & \textbf{100.0} & 99.9          \\
\textbf{Speech commands (V1)}    & 66.9         & 69.4          & 71.7           & \textbf{72.9} \\
\textbf{Speech commands (V2)}    & 44.4         & 54.2          & 62.4           & \textbf{64.2} \\
\textbf{Acoustic scenes}         & 86.4         & 90.7          & \textbf{94.1}  & 90.2          \\
\textbf{Speaker Id. (Voxceleb)}  & 17.6         & 21.6          & \textbf{29.9}  & 22.8          \\
\textbf{Birdsong detection}      & 75.9         & 76.9          & \textbf{77.0}  & 76.4          \\
\textbf{Music, Speech and Noise} & 98.8         & \textbf{99.1} & \textbf{99.1}  & 98.8          \\
\textbf{Language Id.}            & 65.6         & 64.0          & \textbf{71.3}  & 68.4          \\
\textbf{Music instrument}        & 62.3         & 57.3          & \textbf{63.4}  & 56.6          \\ \midrule
\textbf{Average}                 & 68.6         & 70.3          & \textbf{74.3}  & 72.2          \\ \bottomrule
\end{tabular}
\end{table}

\section{Conclusions}
We introduce \ours\!\!, a simple, easy-to-implement, self-supervised contrastive algorithm for general-purpose audio representation learning. Our approach achieves remarkable performance improvements over earlier unsupervised methods on a wide variety of challenging downstream tasks in a linear evaluation protocol as well as significantly improves results over supervised baselines through fine-tuning. We believe that the simplicity of our system, combined with its strong transferability across audio tasks, will pose it as a go-to baseline for future work in self-supervised learning for audio.

\ninept
\bibliographystyle{IEEEbib}
\bibliography{main}
\end{document}